# Fatigue effect on polarization switching dynamics in polycrystalline bulk ferroelectrics

S. Zhukov,[1a] J. Glaum,[2] H. Kungl,[3] E. Sapper,[4] R. Dittmer[4], Y.A. Genenko,[5a] and H. von Seggern[1]

[1]*Institute of Materials Science, Electronic Materials, Technische Universität Darmstadt, Alarich-Weiss-Straße 2, 64287 Darmstadt, Germany*

[2]*Department of Materials Science and Engineering, Norwegian University of Science and Technology, Trondheim, Norway*

[3]*Forschungszentrum Jülich, Institut für Energie- und Klimaforschung IEK-9, 52425 Jülich, Germany*

[4]*Institute of Materials Science, Nonmetallic-Inorganic Materials, Technische Universität Darmstadt, Alarich-Weiss-Straße 2, 64287 Darmstadt, Germany*

[5]*Institut of Materials Science, Materials Modelling, Technische Universität Darmstadt, Jovanka-Bontschits-Straße 2, 64287 Darmstadt, Germany*

## Abstract

Statistical distribution of switching times is a key information necessary to describe the dynamic response of a polycrystalline bulk ferroelectric to an applied electric field. The Inhomogeneous Field Mechanism (IFM) model offers a useful tool which allows extraction of this information from polarization switching measurements over a large time window. In this paper, the model was further developed to account for the presence of non-switchable regions in fatigued materials. Application of the IFM- analysis to bipolar electric cycling induced fatigue process of various lead-based and lead-free ferroelectric ceramics reveals different scenarios of property degradation. Insight is gained into different underlying fatigue mechanisms inherent to the investigated systems.

[a] Authors for correspondence

email: zhukov.tud@gmail.com

and genenko@mm.tu-darmstadt.de

## 1. Introduction

Switching of spontaneous polarization in ferroelectrics in response to an applied electric field is a key process for many applications, most notably, for ferroelectric memories. In perfectly ordered crystals this dynamic property may be described within a classic Kolmogorov-Avrami-Ishibashi (KAI) nucleation and growth theory[1-3] exploiting a single characteristic switching time $\tau$. In ferroelectric materials with spatially random oriented grains like polycrystalline thin films[4-8] or bulk ceramics[9-11] polarization switching becomes dispersive, *i.e.*, exhibits a wide spectrum of switching times, which are associated with different regions in a sample. Statistical distributions of switching times have been discussed as hypothetical concept[4,12] or have been derived based on the concept of diluted defect dipoles.[6,7,13,14] Experimentally, different switching times were recently explicitly resolved within the range of 1 ms - 1 s and even related to certain switching events by *in situ* high-energy X-ray diffraction,[15] the analysis of which seems to be extendable down to the 1 µs range.[16]

In fact, statistical distributions of switching times can even be directly read out from polarization switching measurements over a wide time window if the inhomogeneous field mechanism (IFM) model is applied.[17,18] Experiments of this kind are reported seldom despite presenting a valuable tool for studying dynamic properties[19] alternative to measurements within the frequency domain.[8,20] The IFM- analysis is based on the hypothesis that the local polarization switching occurs according to a step-like process on the logarithmic time scale (in general similar to the KAI- model) with a switching time $\tau(E)$ dependent on the local electric field $E$ utilizing the experimental finding of the strong field dependence of switching dynamics.[21] This approach has proven to work in systems with sufficient randomness in grain orientation, which include all ferroelectric ceramics[17,22-25] and at least some canonical relaxors[18] and organic ferroelectrics.[26,27] It allows the direct extraction of statistical distributions of the local electric field and, consequently, of the local switching times. These characteristics can be related to phase symmetry and microstructure of a particular material.[22] In special cases, it even is possible to obtain microscopic characteristics of the switching process, such as energy and size of a critical domain nucleus in the polarization reversal process.[23-25]

Ferroelectric materials are prone to fatigue, a gradual degradation of material properties, when exposed to mechanical or electrical cycling load. This phenomenon has been carefully investigated since a long time due to its relevance for application oriented material development and has been periodically reviewed.[28-32] Physical mechanisms and apparent characteristics of fatigue depend on chemical composition, structure of materials and physical conditions and the kind of utilized load. A special class of materials are the canonical relaxors that exhibit a transformation between a macroscopically non-polar relaxor state and a polar ferroelectric state upon initial electric poling.[33,34] These compositions experience degradation of their macroscopic parameters similar to classic ferroelectric materials as long as a ferroelectric order exists without electric field application.[35] However, if the phase transformation is reversible and the ferroelectric order decays without electric field application, remarkable fatigue resistance has been observed.[36] The existence of a stable ferroelectric order is as such a prerequisite for the development of distinct fatigue degradation.

Particularly, fatigue has a great impact on the time-dependence of polarization reversal in ferroelectrics – a fact that has never been systematically studied in detail. All ferroelectric ceramics demonstrate loss of saturation and remanent polarizations[28-32] and slowing down of the polarization reversal response[10,11,32] after enduring bipolar electric cycling. However, these properties often can be almost completely rejuvenated by annealing at temperatures above the Curie point.[11] For commercial $Pb(Zr,Ti)O_3$-based compositions the maximum polarization was found to be restored to magnitudes characteristic of the virgin material, whereas the dynamics of the polarization reversal remained considerably slowed down.[17] This means that the switching time distribution is irreversibly modified. The mechanisms behind the irreversible alteration of the switching dynamics are not fully clarified so far. A consistent investigation of the evolution of statistical distributions of switching times as a main quantitative characteristic of dynamic properties in different fatigued materials is essential. This presents the subject of the current study. The structure of the paper is as follows. In Section 2 a general problem of interpretation of statistical distributions of switching times in fatigued materials is discussed and the method of analysis is developed. The general experimental approach and procedures are described in Section 3. Section 4 explains the extended IFM- model step by step on the basis of experimental data obtained from La- doped PZT while Section 5 reports results on a commercial PZT material PIC 151. Section 6 presents a study of Cu- stabilized 94BNT-6BT ceramics. Comparative analysis of different materials and discussion constitute Section 7. Conclusions are given in Section 8.

## 2. IFM- model extension to electrically fatigued materials

For different physical reasons a substantial part of a ferroelectric material becomes unswitchable in the process of electric fatigue.[28-32] This means appearance of numerous regions which, being penetrated by electric field, do not switch even at enhanced field magnitudes and long observation times. Such a behaviour reminds of a composite consisting of dielectric and ferroelectric regions. Description of polarization dynamics for such materials requires modification of the IFM- model which was initially developed for a one-component polycrystalline medium.[17,18] The necessary modifications to the IFM- model are introduced in this section.

The IFM- approach implies that the local polarization switching proceeds according to the KAI- model[1-3] and can be described by the dependence

$$p(t, \tau) = 2P_s \left\{ 1 - \exp(-(t / \tau)^\beta) \right\} \qquad (1)$$

where $P_s$ is the spontaneous polarization, $\beta$ the so called Avrami index which adopts values between 1 and 4 depending on the dimensionality of the reversed polarization domain,[3] $t$ the time elapsed after application of the electric field and $\tau$ the local switching time depending on the local electric field $E$. The latter is understood as coarse-grained, *i.e.* averaged over the individual grain volume. The total polarization response of a virgin polycrystalline bulk ferroelectric is assumed to result from superposition of local responses by independent spatial regions given by Eq. (1):

$$\Delta P(E_m, t) = \langle \cos \theta \rangle_0 \int_0^\infty d\tau \, Q(\tau) \, p(t, \tau) \qquad (2)$$

where $E_m$ is the electric field applied to the ferroelectric, $\langle \cos \theta \rangle_0$ the mean cosine of the polarization polar angle with respect to the applied field direction in the virgin material, and $Q(\tau)$ a weighted statistical distribution function of the switching times.[18,22] Equation (2) generalizes a similar form of the nucleation limited switching (NLS) model suggested for thin polycrystalline ferroelectric films[4] to three-dimensional bulk ceramics.[18] If the field dependence of the local switching time $\tau(E)$ is known, the weighted statistical distribution of times $Q(\tau)$ can be related to the weighted statistical distribution of the reduced field values $f(E / E_m)$ as

$$Q(\tau) = \left| \frac{d\tau}{dE} \right|^{-1} \frac{1}{E_m} f\left( \frac{E}{E_m} \right). \qquad (3)$$

The function $f\left(E/E_m\right)$ can be derived from the full statistical distribution function of the vector field values $F\left(E/E_m,\theta\right)$ as [18]

$$f(s) = \frac{1}{\langle\cos\theta\rangle_0}\int_0^\pi d\theta\sin\theta\cos\theta\,F(s,\theta)\,. \tag{4}$$

where $s$ is a dimensionless variable presenting the reduced electric field $E/E_m$. Both functions $f(s)$ and $Q(\tau)$ are normalized according to their statistical meaning in the virgin material:

$$\int_0^\infty ds\,f(s) = 1 \quad \text{and} \quad \int_0^\infty d\tau\,Q(\tau) = 1\,. \tag{5}$$

The problem arising in the course of electric fatigue is that the polarization in some regions can become unswitchable.[28-32,37] The electric field $E$ remains finite in these regions but the polarization switching is suppressed corresponding effectively to $\tau=\infty$ violating the local dependence $\tau(E)$. Relation (3) is thus violated as well and this circumstance affects the normalization conditions (5). Most easily the behaviour of the two-component composite can be interpreted in terms of characteristic switching frequencies $\omega=1/\tau$. A corresponding statistical distribution function for a fatigued system can then be presented in the form:

$$H(\omega) = h(\omega) + \nu\cdot\delta(\omega) \tag{6}$$

where $h(\omega)$ is a regular part of the distribution while the singular $\delta$-function represents the unswitchable fraction $\nu<1$ of the volume characterized by $\omega=0$ (i.e. $\tau=\infty$). Such a form appears, for example, in the frequency-dependent conductivity of superconductors $\sigma(\omega)$, which obeys the conductivity sum rule, similar to normalization (5), and contains a $\delta(\omega)$-term describing the contribution of superconducting condensate.[38] From Eq. (6), the switching time distribution can be derived in the form:

$$Q(\tau) = \overline{Q}(\tau) + \nu\cdot\frac{1}{\tau^2}\delta\left(\frac{1}{\tau}\right) \tag{7}$$

where $\overline{Q}(\tau)$ describes the regular part of the distribution related to the regions switchable in a finite time. Accordingly, $\overline{Q}(\tau)$ obeys now the normalization condition:

$$\int_0^\infty d\tau\, \overline{Q}(\tau) = 1 - \nu. \tag{8}$$

In the following it is assumed that the unswitchable regions are arbitrarily excluded from the initially switchable sample volume without any correlation with the local electric field. To keep the convenient mapping $E \to \tau$ it is assumed formally that the relation $\tau(E)$ applies in the whole sample volume, then the statistical field distribution splits according to Eqs. (3) and (7) into

$$f(s) = \overline{f}(s) + \nu \cdot \delta(s) \tag{9}$$

as if the value of the field applied to the unswitchable regions is equal to zero. This is followed by the normalization condition for the regular part of the statistical field distribution $\overline{f}(s)$:

$$\int_0^\infty ds\, \overline{f}(s) = 1 - \nu. \tag{10}$$

Thus, in the case of fatigue, $Q(\tau)$ in Eq. (2) should be substituted by $\overline{Q}(\tau)$ which makes a difference to the unfatigued material.[17,18,22] The maximum total polarization then reaches a magnitude:

$$\Delta P_{\max,f} = \Delta P_{\max,0} \frac{\langle \cos\theta \rangle_f}{\langle \cos\theta \rangle_0} \left(1 - \nu\right) \tag{11}$$

where $\Delta P_{\max,0}$ and $\langle \cos\theta \rangle_0$ are the maximum total polarization reversal and the mean cosine of the polarization polar angle in the completely ferroelectric material, respectively, and the corresponding parameters with index $f$ characterize the fatigued material. Thus, the fatigue parameter $\nu$ can be, in principle, evaluated from direct measurements on the virgin and fatigued material using Eq. (11). The mean value of the electric field over the switchable regions is accordingly reduced which results in the approximate normalization condition on $\bar{f}(s)$:

$$\int_0^\infty ds\, s\, \bar{f}(s) \cong \frac{1-\nu}{\langle \cos\theta \rangle_f}. \tag{12}$$

It was experimentally established that, in fatigued PZT bulk ceramics[17] as well as in various ferroelectric ceramics in the virgin state,[18,22-25] logarithmic derivatives of the total polarization with respect to the applied electric field exhibit scaling properties and can be represented as a function of the combined variable of field and time:

$$\frac{1}{\Delta P_{max,0}} E_m \frac{\partial \Delta P}{\partial E_m} = \Phi\left(\frac{E_m}{E_{max}(t)}\right). \tag{13}$$

Here $E_{max}(t)$ is the time-dependent position of the maximum of the derivative in the left-hand side of Eq. (13) presented as the function of field. If this property is satisfied and all derivatives evaluated for different times fall on the same master curve $\Phi(u)$ then the statistical distribution function of fields $\bar{f}(s)$ can be directly related to the master curve as[18,22]

$$\bar{f}(s) = \frac{\langle \cos\theta \rangle_0}{\langle \cos\theta \rangle_f} \frac{1}{s} \Phi\left(\frac{1}{\gamma s}\right). \tag{14}$$

The parameter $\gamma$ can be estimated from the shape of the master curve using Eq. (12) as

$$\gamma \cong \frac{\langle \cos\theta \rangle_0}{1-\nu} \int_0^\infty \frac{du}{u^2} \Phi(u). \tag{15}$$

The dynamic polarization response of a system satisfying the relation (13) reads

$$\Delta P\left(E_m, t\right) = \Delta P_{max,0} \int_0^{E_m/E_{max}(t)} \frac{du}{u} \Phi(u) \tag{16}$$

as in the case of the virgin ceramics. Also the $\tau(E)$ dependence can be extracted from the $E_{max}(t)$ function in the same way.[18,22-25] Detailed explanations of the measured polarization data handling are given in algorithmic form in Ref. [22]. The value of the mean cosines cannot be separately determined from measurements of the total polarization. For simplicity, in the following analysis it will be set to unity. Fortunately, the uncertainty about the parameter $\gamma$ has no effect on the polarization response described by Eq. (16). The latter is determined by the two functions $\Phi(u)$ and $E_{max}(t)$ which present the fingerprints of a bulk polycrystalline ferroelectric in the fatigued state comparable to previous reports on virgin ceramics.[18,22-25] In the following, the IFM- analysis will be applied to three different bulk ferroelectric ceramic compositions subject to bipolar electric cycling.

## 3. Experimental

Three different ceramic compositions were investigated in this study:
1) Pb(Zr$_{52.5}$Ti$_{47.5}$)O$_3$ +2 mol% La abbreviated PZT+La
2) Cu- stabilized 94(Bi$_{1/2}$Na$_{1/2}$)TiO$_3$-6BaTiO$_3$ abbreviated 94BNT-6BT+Cu
3) Pb$_{0.99}$[Zr$_{0.45}$Ti$_{0.47}$(Ni$_{0.33}$Sb$_{0.67}$)$_{0.08}$]O$_3$ abbreviated PIC 151.

PZT+La and 94BNT- 6BT+Cu ceramics were synthesized by a mixed solid state route,[39-41] while PIC 151 is a commercial composition purchased from PI Ceramics (Lederhose, Germany). All the samples were disk-shaped with 6 to10 mm in diameter and 0.5 to 1.0 mm in thickness.

PZT 52.5/47.5 doped with 2 mol% La was synthesized according to Pb$_{0.97}$La$_{0.02}$ (Zr$_{0.525}$Ti$_{0.475}$)O$_3$ stoichiometry following a mixed oxide route.[39,40] PbO, ZrO$_2$, TiO$_2$ and preheated La$_2$O$_3$ were homogenized by attrition milling for 4 h at 1000 rpm in isopropanol. A polyamide crucible, a polyamide stirrer and zirconia (YTZ) milling balls (2 mm in diameter) were used for the milling procedure. After separating the slurry from the milling balls, it was dried in a rotation evaporator and stored for two days under vacuum at 100°C. The dried powder mixture was sieved and then calcined at 850°C for 2 h in air. Subsequent to the calcination, the powder was milled for 6 h at 200 rpm in a planetary ball mill. After drying the powder, it was passed to a 160 µm mesh sieve in order to destroy coarse agglomerates. Dry forming was applied to prepare green bodies of 12 mm in diameter and 10 mm in height by dye pressing and subsequent cold isostatic pressing at 500 MPa. The samples were then sintered at 1050°C for 6 h in closed alumina crucibles under air atmosphere. The sintered bodies had densities of 7.85 g/cm$^3$ and a mean grain size of 1.7 µm. Slices were cut from the inner part of the sintered bodies, then ground and polished to a thickness of approximately 1 mm, before applying silver electrodes by RF- sputtering to the faces of the discs.

94BNT-6BTceramics were prepared using the conventional solid oxide route starting with 94BNT-6BT+Cu with the chemical formula $(Bi_{0.47}Na_{0.47}Ba_{0.06})TiO_3$.[41] High-purity oxides and carbonates (>99.0%, Alfa Aesar, Karlsruhe, Germany), $Bi_2O_3$, $BaCO_3$, $K_2CO_3$, $Na_2CO_3$, and $TiO_2$ were used as starting materials. The powders were weighed according to their stoichiometric formula and ball-milled with yttria-stabilized zirconia balls in an ethanol suspension for 24 h using a planetary mill (Pulverisette 5, Fritsch GmbH, Idar-Oberstein, Germany). The ball-milled suspensions were dried at around 100°C and pulverized afterwards in order to destroy any agglomerates. Calcination was carried out in covered alumina crucibles for 3 h at 900°C with a heating rate of 5°C/min. For one batch, an excess amount of 1.0 mol% CuO was added to the calcined 94BNT-6BT powder. Then, the powders were ball milled again and dried under the same conditions as for the initial milling step. Finally, samples were pressed into cylindrical green pellets. The green bodies were sintered at 1150°C (Nabertherm furnace) with a heating rate of 5°C/min. In order to avoid the loss of volatile components, the samples were embedded in the powder of the same composition during sintering. Sintered samples were ground to a thickness between 350 µm and 380 µm and had a diameter ranging between 7 mm and 7.2 mm. Afterwards the samples were polished using diamond paste with a grain size of 1 µm. To release stresses induced by preparation, the samples were annealed for 2 h at 450°C. Silver electrodes of 50 nm thickness were applied by sputter coating (Emitech K950X). To ensure mechanical stability during cycling, an additional layer of silver paste (Gwent Electronic Materials Ltd, Pontypool, U.K.) was applied and fired at 400°C.

Commercial samples PIC 151 had co-fired silver electrodes (processed at 850°C) applied by the supplier. The samples of all compounds were bipolar fatigued at a field of about a twice the coercive field strength ($E_{fat} \sim 2E_c$) and a frequency of 50 Hz.

For measurements of the switched polarization $\Delta P$ as a function of the applied electric field $E_m$, a voltage pulse was employed for a poling time $t_P$ using an electrical setup and procedure described in Ref. [11]. The applied fields $E_m$ covered the range from $0.3E_C$ to $3E_C$, while the duration of the switching pulse $t_P$ was varied in the wide time domain from 1 µs to 100 s.

To clarify the stability of the induced fatigue degradation the samples were subjected to a high temperature treatment after the final cycling step. For that purpose the fatigued samples were annealed for 24 h in air at $400^0$C which is well above the Curie temperature of $T_C = 250^0$C for PIC 151, $290^0$C for PZT+La and the characteristic temperature $T_m \cong 300$ °C for 94BNT- 6BT+Cu above which the relaxor characteristics start to vanish.[41,42]

## 4. Application of the extended IFM- model to experimental PZT+La data

Figure 1 displays the experimental dependencies obtained for the virgin and fatigued La- doped PZT ferroelectric ceramic for poling pulses ranging from 1 µs to $10^3$ s. As one can see from Fig. 1, the complete switching of polarization strongly depends on the poling time $t_P$ and the applied electric field $E_m$. For example, at a high field of 2.0 kV/mm, the complete

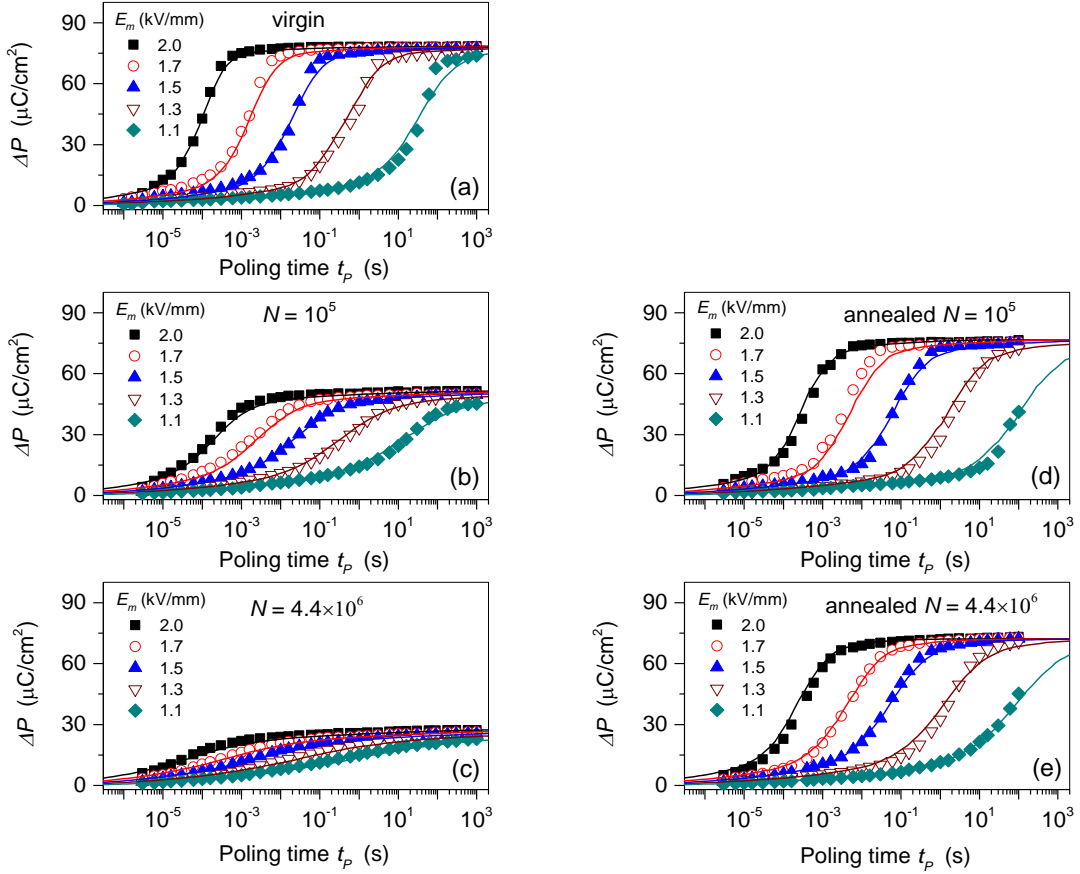

FIG. 1 (color online). Switching dynamics in PZT+La: virgin (a) and after fatigue (b)-(e). Left: (b) $10^5$ and (c) $4.4 \times 10^6$ bipolar cycles; right: (d) and (e) - after annealing of fatigued samples. Symbols correspond to the experimental results while solid lines to the IFM- model results.

switching of the virgin sample (Fig. 1a) already occurs during a short voltage pulse of 1 ms, whereas at 1.1 kV/mm the switching is not completed for pulse lengths $< 10^3$ s. Additionally, with increasing level of fatigue the polarization responses become more dispersive exhibiting a substantially reduced total switched polarization $\Delta P_{\max,f}$. Obtained values for $\Delta P_{\max}$ are presented in Table I and were used to calculate the parameter $\nu$ by Eq. (11).

As was mentioned in Section 2, the polarization switching kinetics in ferroelectrics is traditionally analyzed by the classical KAI- model utilizing Eq. (1). Examples of the best KAI- model fitting to the experimental results measured at $E_m = 1.5$ kV/mm for virgin and fatigued La- doped PZT specimens are displayed in Fig. 2. It is easy to see that the switching responses in both specimens can hardly be fitted by the KAI- model with $\beta = 1$ represented by solid line. A common practice in the case of dispersive polarization response is to allow the parameter $\beta$ in Eq. (1) to take a value less than unity.[10,43] This assumption is questionable because, in the framework of KAI- model, there is no physical meaning for $\beta < 1$ since the dimensionality of a growing nucleus cannot be less than 1. The best KAI fitting with $\beta = 0.58$ (virgin) and $\beta = 0.19$ (fatigued) are also plotted by dashed lines in Figs. (2a) and (2b), respectively. This approximation describes better the switching response for the virgin sample but at the same time it fails to describe the fatigued one. From these fitting examples it is clear that the classical KAI- model is unable to describe the dispersive switching response especially in fatigued ceramics. Therefore, as already mentioned in the Introduction and Section 2 the alternative IFM- approach was utilized to analyze the switching properties of ferroelectric ceramics.

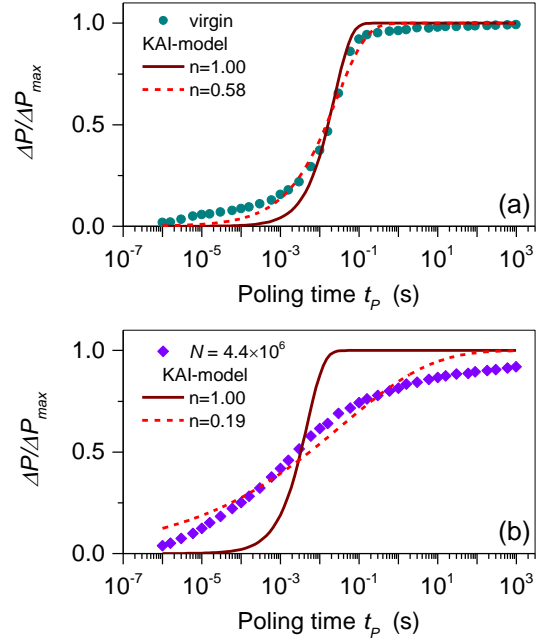

FIG. 2. (color online) Normalized polarization $\Delta P/\Delta P_{max,f}$ versus poling time for PZT+La (a) virgin and (b) fatigued specimens measured at $E_m = 1.5$ kV/mm and room temperature. Solid and dashed lines represent KAI- model fits with two different $\beta$ values as indicated.

In the following it will be demonstrated how the two functions $\Phi\left(E_m / E_{\max}(t)\right)$ and $E_{\max}(t)$ utilized in the calculation of the total switching response by Eq. (16), are obtained from the experimental results depicted in Fig. 1. This will be explained in detail for a virgin sample. To this end, the field dependence of $\Delta P$ for fixed poling times $t_P$ is re-plotted as a function of $E_m$, as shown in Fig. 3(a). Then, the logarithmic derivatives for each $t_P$ are calculated and normalized by $\Delta P_{\max,0}$ as follows:

$$\frac{1}{\Delta P_{\max,0}} \times \frac{dP}{d\left(\ln E_m\right)} = \frac{1}{\Delta P_{\max,0}} \times \frac{dP}{dE_m}\left(\frac{d\left(\ln E_m\right)}{dE_m}\right)^{-1} = \frac{E_m}{\Delta P_{\max,0}} \times \frac{dP}{dE_m}, \qquad (17)$$

$\Delta P_{\max,0} = 78.5 \ \mu C/cm^2$ was obtained from Fig. 3(a). Then it is assumed that after long poling times all $\Delta P\left(E_m\right)$ curves presented in Fig. 3(a) reach the same maximum value of $78.5 \ \mu C/cm^2$. Resulting normalized logarithmic derivatives for different poling times $t_P$ are plotted in Fig. 3(b).

It can be seen that the amplitudes of the maxima in Fig. 3(b) are practically independent of the actual value of poling time which is not the case for single crystal samples.[18,25] This indicates that the scaling of all curves to the same shape is in principal possible. Since the data points are scattered the data sets for certain poling times $t_P$ were first spline-fitted and then the maximum position $E_{\max}$ was determined from the maximum of the fitted curve. The scaling of $E_m$ values of each curve in Fig. 3(b) to its own maximum position $E_{\max}$ results in a master curve $\Phi\left(E_m / E_{\max}(t)\right)$ as displayed in Fig. 3(c). The existence of such a master curve is the precondition of the applicability of the IFM- model.[17,18]

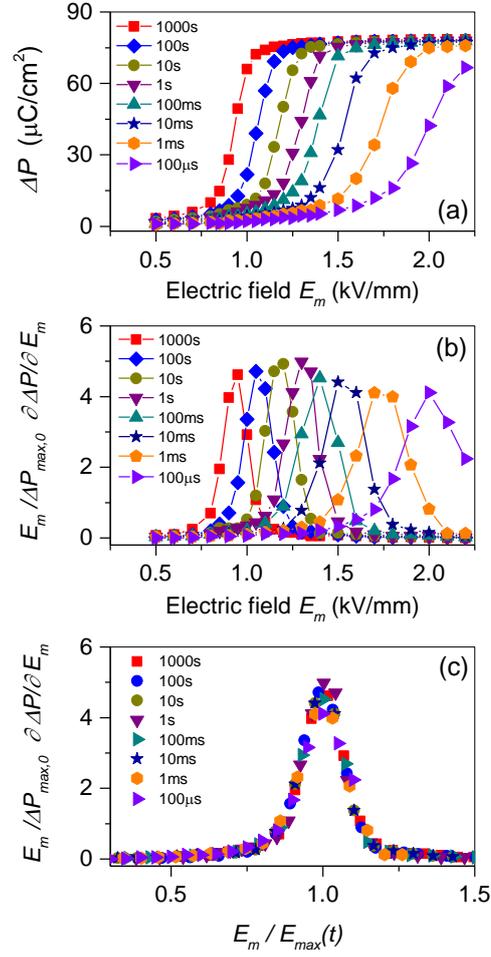

FIG. 3 (color online). Switched polarization $\varDelta P$ of virgin PZT+La ceramic versus applied field $E_{\mathrm{m}}$ at different poling times $t_P$ as indicated (a), its normalized logarithmic derivatives versus applied field (b), the same derivatives scaled to their maximum positions $E_{\max}(t)$ (c).

So obtained master curves $\Phi$ for virgin and differently fatigued samples were used to restore the distribution function for local fields $f\left(E/E_m\right)$ utilizing Eq. (14). The master curves and corresponding $f\left(E/E_m\right)$ distributions are shown in Fig. 4. One recognizes that the obtained master curves and local field distributions change continuously with the number of $ac$ load cycles $N$, showing rather asymmetrical and extremely broad distributions for higher cycle numbers.

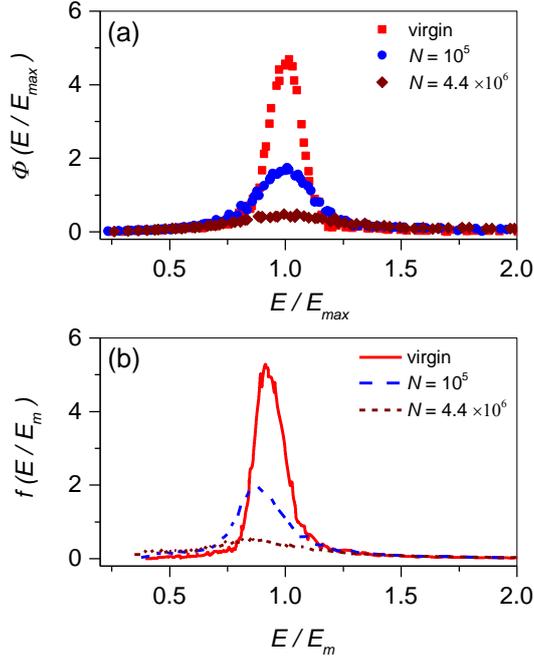

FIG. 4 (color online). (a) Master curve $\Phi\left(E_m / E_{max}(t)\right)$ and (b) local field distribution $f\left(E / E_m\right)$ for differently fatigued PZT+La samples as indicated.

Prior to applying Eq. (16) to the calculation of the polarization reversal it is useful (though not necessary) to analytically fit the dependence of $E_{max}$ on the poling time. The experimentally determined $E_{max}(t)$ reflects the actual field dependence of the local switching times $\tau(E)$, where $E$ represents the local electric field. Different empirical and theoretical expressions for $\tau(E)$ have been proposed to relate the switching time to the applied electric field.[21,44-46] At this point it should be noted that the IFM- model does allow for a free choice of the mechanism for the reverse domain nucleation and growth, and does not assume *a priori* a certain microscopic switching scenario. The experimental function $E_{max}(t)$ can, however, help to identify the suitable theoretical mechanism. According to the IFM- analysis the local switching time $\tau(E)$ results from solving the equation $E_{max}(t) / \gamma = E$ with respect to *t*. The parameter $\gamma$ involved here is calculated from the shape of the master curve $\Phi$ (see Eq. (15)) and typically adopts values around unity.

The equation above can easily be solved numerically by simply exchanging the $E$ and $t$ axes in the dependence $E_{max}(t)$ and consequent scaling with the factor $\gamma$ on the field axis. The obtained experimental data for the PZT+La composition and resulting fits by the empirical Merz function[21]:

$$\tau(E) = \tau_0 \times \exp\left[\left(\frac{E_a}{E}\right)^{\alpha}\right], \tag{18}$$

are displayed in Fig. 5. $\tau_0$ is the characteristic time constant, $E_a$ the activation field and $\alpha$ an additional fitting parameter. The shape of the dependence $\tau(E)$ and the parameters extracted by fitting allows one to judge on the microscopic mechanism of the reversed domain nucleation and the activation barrier height. To improve the stability of the fitting procedure by Eq. (18), the parameter $\alpha$ was set to unity for all the ceramics investigated. It can be seen in Fig. 5, that the selected function provides a good fit for the whole time interval ranging over 8 decades for the virgin and differently fatigued ceramics. Obtained parameters $\gamma$, $\tau_0$ and $E_a$ are summarized in Tab. I.

Knowing the functions $\Phi\left(E_m / E_{max}(t)\right)$ and $E_{max}(t)$ is sufficient to describe the macroscopic polarization response of the system by Eq. (16). Obtained results for different applied field strength are displayed in Fig. 1 by solid lines. In all cases, a good agreement between the experimental data and the IFM- model calculations was achieved.

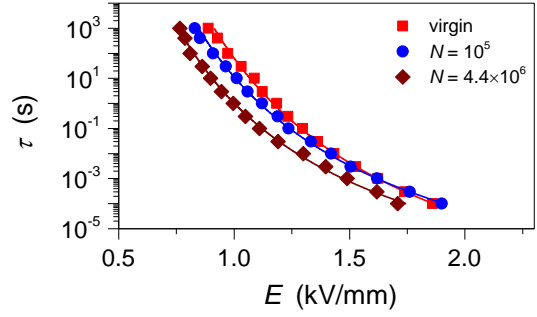

FIG. 5. $\tau$ versus local field $E$ for differently fatigued PZT+La specimens as indicated. Symbols correspond to experimental results while solid lines represent the fits by Eq. (18).

TABLE I. The IFM- model parameters for PZT+La ceramics.

| Cycle number $N$ | $\Delta P_{max}$ ($\mu$C cm$^{-2}$) | $E_a$ (kV mm$^{-1}$) | $v$ | $\tau_0$ (s) | $\gamma$ |
|---|---|---|---|---|---|
| Virgin | 78.4 | 28.8±0.3 | 0.00 | (1.9±0.4)×10$^{-11}$ | 1.076 |
| 10$^5$ | 51.5 | 24.4±0.3 | 0.34 | (2.8±0.5)×10$^{-10}$ | 1.125 |
| Annealed 10$^5$ | 76.5 | 30.0±0.6 | 0.02 | (3.4±1.2)×10$^{-11}$ | 1.059 |
| 4.4×10$^6$ | 27.5 | 21.6±0.4 | 0.65 | (4.2±1.2)×10$^{-10}$ | 1.119 |
| Annealed 4.4×10$^6$ | 72.2 | 29.9±0.3 | 0.08 | (2.9±0.5)×10$^{-11}$ | 1.067 |

One of the major objectives of the present study is to establish the evolution of the statistical distribution of switching times $Q(\tau)$ with fatigue. The knowledge of the two dependencies $f\left(E/E_m\right)$ and $\tau(E)$ allows for the determination of $Q(\tau)$ by Eq. (3). The switching time distributions of differently fatigued samples for the applied field of 1.5 kV/mm are exemplarily displayed in Fig. 6(a) in logarithmic representation as $G(\ln(\tau/\tau_0))=\tau Q(\tau)$. The distribution of switching times is much broader for the fatigued specimens than for the virgin

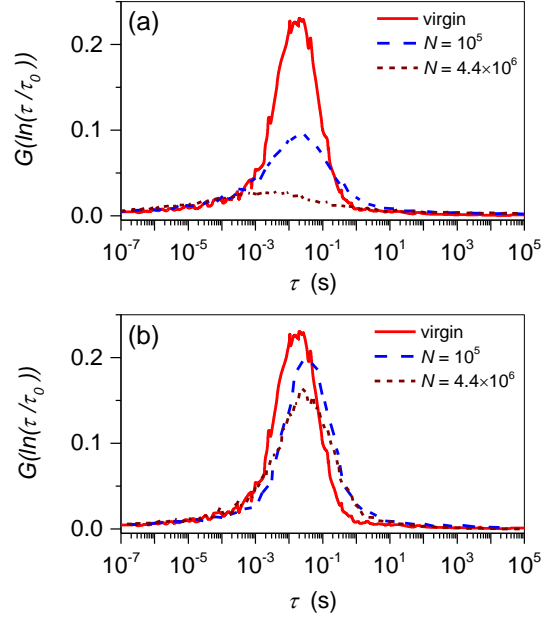

FIG. 6 (color online). Switching time distributions at $E_m = 1.5$ kV/mm for PZT+La ceramic with different level of fatigue as indicated: (a) after fatigue and (b) after fatigue and consecutive annealing for 24 h in air at $400^0$C.

one. Furthermore, the maximum of $G(\ln(\tau/\tau_0))$ is shifted towards faster times for heavily fatigued samples. The latter result is rather unexpected since for most fatigued ferroelectrics a drastic decrease in switching rates was observed.[10,32] On the other hand, the distribution is hardly affected by fatigue for very short switching times.

To discriminate different possible fatigue mechanisms the impact of thermal annealing on polarization switching dynamics is further considered which is known to have a strong effect on fatigued ceramics.[10,11] Figure 1 compares polarization switching responses for fatigued specimens before (left-hand side) and after (right-hand side) annealing. The parameters: $\Delta P_{max}$, $\gamma$, $\tau_0$ and $E_a$ for annealed specimens are listed also in Tab. I, while the field-related calculations by the IFM- model are presented by solid lines in Fig. 1. In general, a significant recovery of the temporal polarization behavior after thermal treatment could be achieved. This is also confirmed by the switching time spectra for annealed specimens presented in Fig. 6(b) which, however, remain somewhat retarded with respect to the virgin material.

## 5. Evolution of dynamic properties in commercial PZT (PIC 151)

The fatigue behaviour of PIC 151 ceramics has been widely investigated in literature.[10,11,17,47-50] Moreover, in a previous study by the present authors the polarization responses in differently fatigued specimens by means of the IFM- model were already analyzed.[17] The analysis presented here is based, however, on the extended version of the IFM- model, as introduced in Section 2.

As an example, Fig. 7(a) displays the temporal evolution of switched polarization for differently fatigued PIC 151 ceramics at an applied field of 1.5 kV/mm. Experimental results

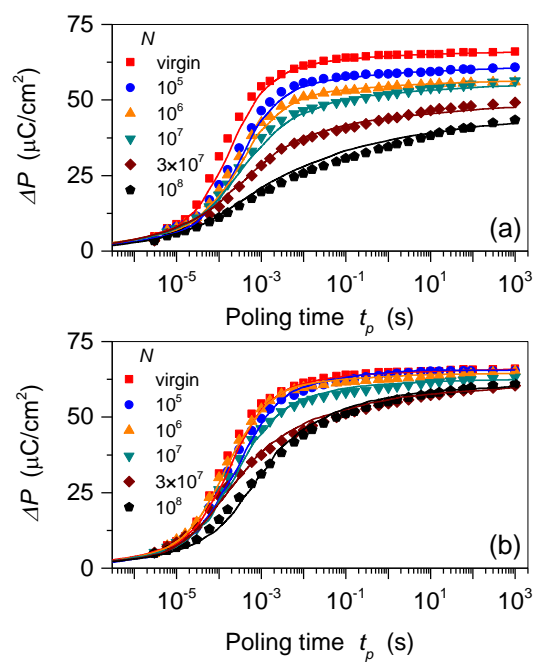

FIG. 7 (color online). Switching dynamics at $E_m$ = 1.5 kV/mm in (a) virgin and fatigued PIC 151 and (b) fatigued ceramics after annealing for different cycles numbers $N$ as indicated. Symbols correspond to the experimental results while solid lines to the IFM- model calculations.

are shown by symbols and corresponding IFM- model calculations by solid lines which provide a good description of the polarization response in time interval of nine orders of magnitude. All the polarization responses depicted in Fig. 7(a) are characterized by fast and slow components. The fast component is realized in the time interval between $10^{-6}$ s and $10^{-2}$ s while the slow component developed for longer times where the polarization practically exhibits a linear dependence on logarithmic time scale.

The impact of thermal annealing on polarization switching dynamics of previously fatigued PIC 151 ceramics was studied as well. Figure 7(b) compares polarization switching responses of a virgin sample with the fatigued ceramics after annealing utilizing the same applied field of 1.5 kV/mm. It was found that for cycle numbers $N \leq 10^6$ a full recovery of the temporal polarization behavior after thermal annealing was obtained. For cycling durations of $10^7$ and larger only a partial recovery of the total polarization was possible, but the poling times had to be prolonged up to $10^3$ s. These samples still display a slow quasi-logarithmic component of polarization reversal as was observed before annealing in Fig. 7(a). The explanation for this phenomenon is that thermal annealing cannot close the micro-cracks generated during cycling which result in a broader field distribution inside the ferroelectric as was discussed previously.[17] At the same time, the loss in switchable polarization observed at the initial stages of fatigue can be explained by charged defects or defect dipoles acting as domain wall pinning centers[28-32] which can be completely removed or redistributed by the heat treatment at temperatures above $T_C$. The presented results suggest that the micro-cracks themselves cannot drastically suppress the switchable polarization but can modify the local field distribution thus causing the changes in the polarization reversal dynamics, *i.e.*, considerably delay the complete switching. Simulations have indicated that cracks, which extend perpendicularly to the electric field direction, shift the statistical field distributions to lower field values[17] thus promoting longer switching times.

TABLE II. The IFM- model parameters for PIC 151 ceramics.

| Cycle number N | $\Delta P_{max}$ ($\mu$C cm$^{-2}$) | $E_a$ (kV mm$^{-1}$) | $\nu$ | $\tau_0$ (s) | $\gamma$ |
|---|---|---|---|---|---|
| virgin | 66.4±0.3 | 24.6±0.6 | 0.00 | $(3.5\pm0.5)\times10^{-12}$ | 1.043 |
| $10^5$ | 61.9±0.3 | 24.6±0.7 | 0.07 | $(5.9\pm0.7)\times10^{-12}$ | 1.043 |
| Annealed $10^5$ | 66.2±0.3 | 28.1±0.5 | 0.00 | $(3.0\pm0.5)\times10^{-12}$ | 0.982 |
| $10^6$ | 57.5±0.2 | 25.4±0.6 | 0.13 | $(4.7\pm0.5)\times10^{-12}$ | 1.054 |
| Annealed $10^6$ | 65.2±0.3 | 27.4±0.6 | 0.02 | $(2.5\pm0.6)\times10^{-12}$ | 0.988 |
| $10^7$ | 56.0±0.2 | 28.2±0.6 | 0.16 | $(4.4\pm0.5)\times10^{-12}$ | 0.987 |
| Annealed $10^7$ | 63.1±0.3 | 28.2±0.5 | 0.05 | $(2.3\pm0.9)\times10^{-12}$ | 0.987 |
| $3\times10^7$ | 52.1±0.2 | 29.3±0.7 | 0.22 | $(5.7\pm0.6)\times10^{-12}$ | 0.958 |
| Annealed $3\times10^7$ | 60.8±0.3 | 28.3±0.5 | 0.08 | $(3.1\pm0.9)\times10^{-12}$ | 0.966 |
| $10^8$ | 48.4±0.2 | 29.5±0.5 | 0.27 | $(1.1\pm0.2)\times10^{-11}$ | 0.940 |
| Annealed $10^8$ | 61.3±0.3 | 29.5±0.6 | 0.08 | $(5.1\pm0.6)\times10^{-12}$ | 0.969 |

Since the main purpose of the current research is to investigate the evolution of the switching dynamics during fatigue, it will be demonstrated in the following how the above-mentioned mechanisms affect the switching time distribution. Like in the case for La- doped PZT, the two dependencies $f\left(E/E_m\right)$ and $\tau\left(E\right)$ have to be determined prior to calculation of $Q(\tau)$ by Eq. (3). Both functions $f\left(E/E_m\right)$ and $\tau\left(E\right)$ together with other switching relevant parameters summarized in Tab. II were obtained in previous work.[17] Accordingly, Fig. 8(a) depicts the calculated switching time distributions for PIC 151 ceramic subjected to persistent bipolar electrical fatigue of different durations and Fig. 8(b) the respective distributions after annealing.

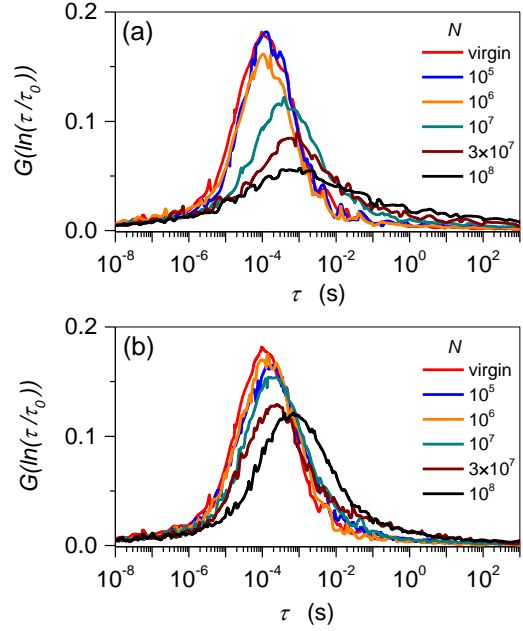

FIG. 8 (color online). Switching time distributions at $E_m$= 1.5 kV/mm for PIC 151 ceramic with different levels of fatigue as indicated: (a) immediately following fatigue and (b) after fatigue and consecutive annealing for 24 h in air at 400$^0$C.

It can be seen in Fig. 8(a) that up to $N \leq 10^6$ cycles the switching time distributions remain qualitatively of the same shape and same maximum position as the distribution for the virgin material. With increasing number of cycles the samples exhibit remarkably asymmetrical distributions in favour of larger switching times. In the previous work scanning electron microscope (SEM) images taken for differently fatigued PIC 151 ceramics were used to reveal the formation of cracks.[11] Almost no cracks were found for samples fatigued up to $N = 10^6$ cycles. For cycle numbers $N \geq 10^7$ the micro-crack formation starts to become notable. No indications of a "dead layer" underneath the electrodes were detected for all samples investigated as was reported previously. Therefore, it is reasonable to propose that formation of mesoscopic cracks of certain shape and orientation can result in an asymmetric distribution of the local electric fields $f\left(E/E_m\right)$ and corresponding switching time distribution.[17] Additionally, the amount of switched polarization $\Delta P_{\max,f}$ gradually decreases with increasing number of cycles in comparison to the virgin sample. The microscopic parameters $\tau_0$ and $E_a$, however, turn out to be less sensitive to the level of fatigue than in the La- doped material (see Tab. I and Tab. II). Annealing recovers the switching time distributions well for cycle numbers up to $N = 10^6$ to $10^7$ as is seen in Fig. 8(b). For higher cycle numbers the switching time spectra are partly recovered but remain essentially dispersive and shifted to higher switching times.

Concluding for PIC 151, the short-time processes are less modified by fatigue than in the above presented La- doped PZT, whereas the long-time processes are affected even stronger. Being more fatigue resistant, PIC 151, however, allows only the restoration of the total polarization by annealing while the dynamics of heavily fatigued materials remains significantly retarded.

## 6. Evolution of dynamic properties in Cu- stabilized BNT-BT

Among the lead-free piezoelectric ceramics developed in the last two decades[31,51] composition $0.94Bi_{0.5}Na_{0.5}TiO3–0.06BaTiO_3$ (94BNT-6BT) shows particularly good piezoelectric characteristics. This composition differs from the PZT- based materials discussed above in the sense that its average crystallographic structure appears to be cubic in the unpoled state. Furthermore, a phase transformation into a polar state occurs upon first application of an electric field,[34] and ferroelectric characteristics such as a distinct domain structure[52,53] and a strain hysteresis loop[54] are observed. These characteristics combined with the diffuse and dispersive appearance of the temperature-dependent permittivity indicate that this material is a canonical relaxor.[55] The stability of the induced ferroelectric phase depends significantly on the transition temperature between the ergodic and non-ergodic relaxor state. For the composition 94BNT-6BT it was determined to be 83ºC[41] – high enough to allow the development of a stable ferroelectric long-range order at room temperature. This is a distinct difference to other BNT-BT based materials for which the transition temperature is lower such that those reveal reversible field-induced phase transitions.[54,56] There are several reports on the bipolar electric fatigue in 94BNT-6BT ceramics, which in general demonstrate a rather low fatigue-resistance of this material.[31,35,57] A strong degradation of macroscopic electromechanical properties is already observed within the first 100 bipolar cycles.[57] Additionally, the domain morphology is strongly affected by long-term cycling showing drastic fragmentation of the domains down to the nano-scale.[58,59] However, the addition of 1 mol% CuO successfully stabilizes the fatigue-resistant tetragonal phase at least for 100 cycles and retains the initial electromechanical properties without significant change.[57] Such a promising finding made the Cu- stabilized 94BNT-6BT ceramic interesting for further investigations of the long-term fatigue resistance. Therefore, in the current study, contrary to the previous reports, the number of bipolar cycles $N$ was remarkably increased reaching a value of $2{\times}10^8$.

Figure 9 shows polarization switching for Cu- stabilized 94BNT-6BT ceramic recorded at an applied electric field of $E_m = 4.0$ kV/mm for a virgin and fatigued specimens as indicated. Almost no changes in switching dynamics were observed up to $10^6$ bipolar cycles. Starting at $N = 10^7$ cycles the switching response is substantially retarded in comparison to the virgin sample, however, the amount of switched polarization $\Delta P_{max,f}$ remains nearly the same. Most dramatic changes were observed for cycling durations between $10^7$ and $10^8$, where a sharp drop of $\Delta P_{max,f}$ is followed by extremely dispersive polarization switching. Annealing after $10^8$ cycles the sample exhibits notable recovery of the total polarization, however, the dispersive quasi-logarithmic response behaviour remains.

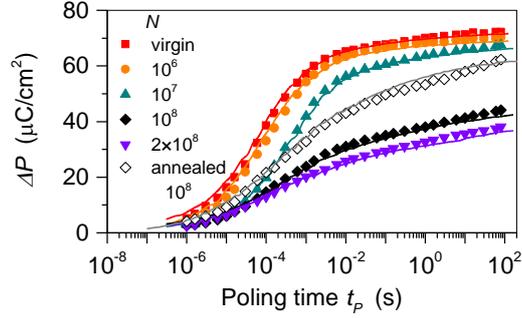

FIG. 9 (color online). Switching response at $E_m = 4.0$ kV/mm in virgin and fatigued 94BNT-6BT+Cu ceramic as indicated as well as in an annealed sample. Symbols correspond to the experimental results while solid lines display the IFM- model calculations.

The IFM- model analysis applied to the 94BNT-6BT+Cu ceramic provides a good description of experimental results for virgin and differently fatigued specimens as shown by the solid lines in Fig. 9, and allows for the determination of the switching time distributions which are depicted in Fig. 10. Switching relevant parameters like $\Delta P_{max}$, $\gamma$, $\tau_0$ and $E_a$ can be found in Table III. It follows from Fig. 10 that almost no changes are observed in the switching time spectra up to $N = 10^6$. For $N = 10^7$ the maximum of $G(\ln(\tau / \tau_0))$ is shifted towards longer switching time constants, however, its amplitude seems to be unchanged with respect to the virgin material. Further increase of the number of cycles leads to an extremely broad distribution of switching times. At the same time the activation field $E_a$ shows a clear trend to higher values with increasing $N$, while the characteristic switching time $\tau_0$ remains practically unaffected as can be seen in Tab. III. Another striking feature is that the specimen fatigued to $N = 10^8$ cycles shows only 33% loss in switchable polarization $\Delta P_{max}$ (see the parameter $\nu$ in Tab. III). This result is comparable to the commercial PIC 151 ceramic subjected to the number of cycles (see Tab. II). One can conclude on the basis of this result that the lead-free 94BNT-6BT+Cu ceramic shows much better long-term fatigue resistance than the La- doped PZT presented in Section 4 and approaches the performance of the commercial PZT (PIC 151) studied in Section 5. On the other hand, the response of the fatigued samples is much more dispersive and retarded than that of PIC 151 (see Fig. 8(b) and 10(b)).

TABLE III. The IFM- model parameters for 94BNT-6BT+Cu ceramics.

| Cycle number N | $\Delta P_{\max}$ ($\mu$C cm$^{-2}$) | $E_a$ (kV mm$^{-1}$) | $v$ | $\tau_0$ (s) | $\gamma$ |
|---|---|---|---|---|---|
| virgin | 75.7±0.4 | 55.9±0.5 | 0.00 | $(5.0\pm0.7)\times10^{-11}$ | 1.030 |
| $10^6$ | 72.0±0.4 | 56.2±0.7 | 0.05 | $(7.7\pm0.9)\times10^{-11}$ | 1.009 |
| $10^7$ | 70.9±0.3 | 64.3±0.9 | 0.06 | $(4.4\pm0.8)\times10^{-11}$ | 0.974 |
| $10^8$ | 50.5±0.2 | 73.9±1.1 | 0.33 | $(2.9\pm1.1)\times10^{-11}$ | 0.875 |
| Annealed $10^8$ | 64.5±0.3 | 66.3±0.9 | 0.15 | $(9.3\pm1.1)\times10^{-11}$ | 0.943 |
| $2\times10^8$ | 41.5±0.1 | 68.3±0.9 | 0.45 | $(4.8\pm1.4)\times10^{-11}$ | 0.923 |

The thermal annealing effect on polarization switching dynamics in fatigued Cu- stabilized 94BNT-6BT ceramic demonstrates unexpected features. Figure 10(b) compares $G(\ln(\tau/\tau_0))$ distributions for a virgin sample of the ceramic and fatigued by $N = 10^8$ cycles, before and after annealing. One can see from Tab. III and Figs. 9 and 10(b) that, on the one hand, annealing partially recovers the amplitude of $\Delta P_{\max}$, but, on the other hand, the switching response remains rather dispersive revealing a broad spectrum of switching times.

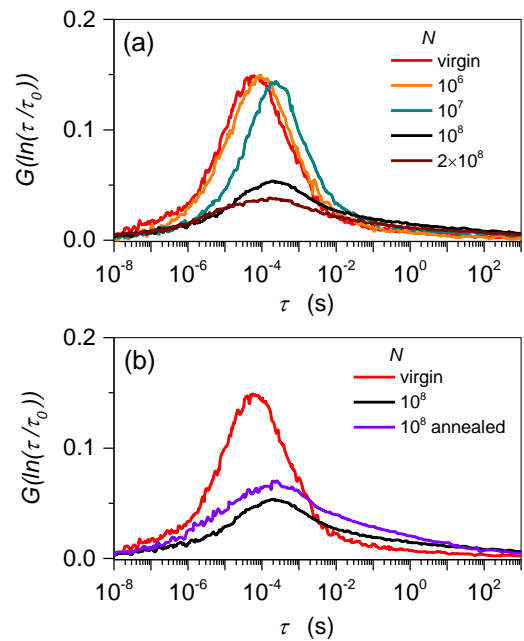

FIG. 10 (color online). Switching time distribution calculated at $E_m = 4.0$ kV/mm for (a) virgin and fatigued 94BNT-6BT+Cu specimens with different level of fatigue $N$ and (b) specimens with $N = 10^8$ right after fatigue and after fatigue with consecutive annealing for 24 h in air at $400^0$C as indicated.

This behaviour with respect to annealing of the fatigued material is similar to that of PIC 151. Different is that the fast switching stage in PIC 151 is hardly affected during fatigue while in 94BNT-6BT+Cu ceramic it is substantially retarded during fatigue and not recoverable by annealing. Unfortunately, no TEM or SEM data on possible micro-cracking are available.

## 7. Discussion

The present study is focused on the investigation of the evolution of statistical distributions of switching times and related key microscopic parameters of polarization reversal in ferroelectric ceramics subjected to prolonged bipolar fatigue. It is evident from the experimental results presented in Sections 4 to 6 that the bipolar fatigue entails a substantial variation in polarization switching dynamics. Figure 11 depicts the evolution of the main polarization reversal characteristics, such as $\nu$, $\tau_0$ and $E_a$ with fatigue for all studied ceramics. Note that the parameter $\nu$ in the framework of the advanced IFM- model represents the volume fraction of the unpolarizable material damaged in the course of fatigue. Polarization in the damaged regions does not switch anymore even at enhanced electric field magnitudes up to $E_m = 2E_c$ applied for a long time. Two other parameters, $\tau_0$ and $E_a$, are obtained from the fit of the experimentally determined dependence of the local switching time $\tau(E)$ utilizing Eq. (18).

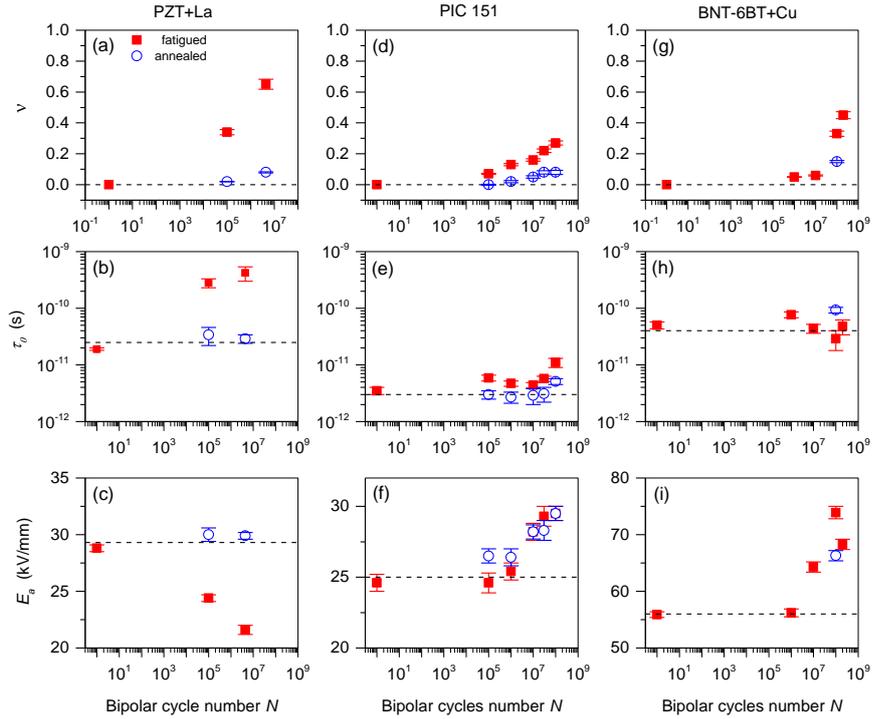

FIG. 11 (color online). Polarization switching parameters extracted from the dynamic measurements for the PZT+La (a)-(c), PIC 151 (d)-(f) and 94BNT-6BT+Cu (g)-(i) ceramics versus the number of bipolar cycles $N$. Solid symbols correspond to the fatigued samples while the open to the samples subjected to thermal treatment for 24 h after fatigue.

One can see in Fig. 11 that the La- doped PZT ceramic reveals the weakest fatigue resistance among all tested materials. During the first $10^5$ cycles already up to 40% of the sample volume are damaged. Further fatigue additionally enhances the effect. The microscopic parameters $\tau_0$ and $E_a$ seem also remarkably dependent on fatigue but exhibit controversial trends. Thus, the activation field $E_a$ decreases with fatigue in the La- doped PZT but increases in PIC 151 and 94BNT-6BT+Cu. Besides, as was mentioned in Section 4, fatigue drastically changes the distribution of switching times (as can be compared in Figs. 6(a), 8(a) and 10(a)) making it much broader and affecting the longer switching times first. At the same time, the left part of the spectrum showing the shorter switching times (and higher electric fields in corresponding regions) remains unchanged with fatigue in La- doped PZT and PIC 151 ceramics. Evolution of the switching time distribution in 94BNT-6BT+Cu reveals more complicated traits. It does not change at all up to $10^6$ cycles, then undergoes a remarkable shift towards longer switching times at $10^7$ cycles and afterwards follows the general behavior of fatigued PZT. From the point of view of the IFM- model, assuming a unique switching mechanism of polarization reversal at all locations in the system, the behavior of PZT materials indicates that fatigue starts to influence initially the areas of material where the local electric fields are weak in comparison with the mean applied field $E_m$. This assumption is, however, not necessarily true as some direct observations prove.[15] The elementary 180°- and non-180°-switching processes exhibit different characteristic switching times and are related to different energy barriers $U_a$. Comparing the Merz empirical law,[21] Eq. (18), with the concept of thermally activated switching,[44-46] $\tau(E) = \tau_0 \times \exp\left(U_a / kT\right)$, it is apparent that distinct switching processes are expected to be characterized by distinct activation fields $E_a$. Different mechanisms of polarization reversal may also be differently affected by fatigue. Particularly, the decrease of the activation field $E_a$ in the La- doped PZT (see Fig. 11(c)) may indicate that processes with lower switching rates are completely blocked in the course of fatigue in contrast to PIC 151 and 94BNT- 6BT+Cu (Figs. 11(f)-(i)). Keeping this in mind, in the following particular features of evolution of the considered materials in the course of fatigue were analyzed.

Investigations on fatigue behavior of La- doped PZT were reported since the 1990s, when changes in polarization and coercive field strength alternations with the number of switching cycles in dependence on materials characteristics, electrode preparation and field loading were analyzed.[60-63] These experiments were focused on materials with high Lanthanum content (PLZT La/Zr/Ti 7/68/32), rhombohedral structure, low lattice distortion and on sample preparation employing sputtered gold electrodes. The influences of grain size, porosity and surface morphology were examined and it was found that large grain size as well as high porosity accelerates the fatigue, indicated by decrease in switchable polarization and increase in coercive field. Grain size dependent crack formation was considered to be the major issue in the fatigue process.

Analysis of fatigue behavior of PZT+La ceramics from the tetragonal side close to the morphotropic phase boundary (Zr/Ti ratios of 52.5/47.5) for PZT doped with 2 mol% La, 1 mol% La, and 1 mol% La +0.5 mol% Fe further elucidated the effects of damage in the surface-near regions under electrode preparation with silver pastes burned in at 400°C.[64] Along with this type of electrode preparation pronounced fatigue was detected already at $3 \times 10^4$ cycles. Electrode near surface crack formation was found to be the main reason for the degradation of ferroelectric properties. Removal of the surface-near layer almost completely recovered the initial performance, so no significant fatigue in the bulk could be detected. It was concluded, that the formation of the cracks in the surface layer leads to screening of the bulk from high fields and therefore prevents electrical fatigue in the bulk. Progress in fatigue scaled with La- donor content (or excess) and grain size. High donor content and low grain size are reducing the fatigue in these materials.

Two aspects will now be addressed based on the present experiments and the IFM model evaluation: (i) the fatigue related to the distribution of switching times and (ii) the fast fatigue of PZT+La without crack formation.

(i) The fatigue behavior indicates that dominant fatigue occurs in particular for the slow polarization switching processes, whereas the contribution of the fast switching processes with $\tau < 10^{-3}$ s remains almost the same up to $10^5$ cycles (Fig. 6). A key to understand this different fatigue behavior depending on switching times is to distinguish between 180° and non 180° switching processes. Owing to the changes in mechanical stress involved in non-180° processes, this type of switching will be slower than that of a one-step 180° process. Therefore, one might attribute the slow processes subject to pronounced fatigue to non-180° switching, whereas the fast polarization reversal processes not subject to fatigue reflect the 180° switching. Although this line of argument seems to be convincing, a more detailed discussion of this issue is necessary. First, one would expect the fraction of 180° switching events contributing the total polarization to be quite high, whereas the fraction of the "fast" processes subject to fatigue seems to be limited as seen from the experiments. Second, the open question remains[15] if (i) a 180° switching event proceeds actually as a one-step process or (ii) is the result of two consecutive non-180° switching events or (iii) both switching mechanisms have to be considered. Under the assumption that in PZT+La three types of switching processes occur – non-180° switching, one-step 180° switching and 180° switching with two consecutive non-180° steps, the conclusion with respect to a description of the fatigue behavior has to be specified as follows: the pronounced fatigue in the slow processes can be attributed to the non-180° switching and the two-step 180° switching, whereas the fast one-step 180° processes are not subject to fatigue up to $10^5$ cycles.

(ii) The absence of micro-crack-induced fatigue as evidenced by the almost complete restoring of the switchable polarization and its dynamics by heat treatment (Fig. 1) seems contrary to what is expected when comparing the present results on PZT+La with those on PIC 151 and with previous work on PZT+La. The higher lattice distortion of the predominantly tetragonal structured PZT+La could lead to locally higher stresses from non-180° switching and consequently enhanced formation of micro-cracks. However, the lattice distortion and the porosity is not the only factor of influence on the micro-crack formation. The most important characteristic of the PZT+La ceramics, which may reduce the crack formation compared PIC 151 is the smaller grain size of 1.7 µm for PZT+La (the same as in Ref. [64]) compared to 6 µm for the PIC 151.[10] The critical grain size leading to spontaneous crack formation was calculated to be 7.05 µm for strain energies and fracture surface energies for morphotropic PZT ceramics.[60] Therefore from theoretical point of view it seems possible that the absence of the micro-crack formation in the PZT+La can be attributed to the relatively small grain size in the PZT+La ceramics. On the other hand, substantial fatigue initiated by the formation of micro-cracks in the surface layer of PZT + 2% La ceramics with the same grain size and lattice distortion has been observed.[64] However, with the micro-cracking occurring in the interface region of the ceramics and the electrode, the stress imposed by the electrode strongly modifies the stress within the material at the interface. Therefore, the differing behavior between PZT + 2% La in previous[64] and in the present experiments could be due to different electrode preparation procedures. For the fatigue mechanism in the PZT+La it is concluded that the small grain size prevents crack formation, thus giving rise to switching under an unscreened field which leads to fast fatigue in this material involving charged defects.[32]

In comparison to the La- doped PZT, the commercial PZT ceramic PIC 151 shows much better fatigue resistance. Almost no changes in switching characteristics were observed for cycle numbers of up to $N \leq 10^6$ (see Figs. 7, 8(a) and 11(d)-(f)). Based on the IFM- model analysis of the polarization responses in differently fatigued specimens, two main mechanisms of degradation can be suggested. The first one, dominant at low cycle numbers up to $10^7$, causes a minor reduction of the switchable polarization without qualitative changes to the distribution of switching times (see Fig. 8(a)). Pinning of domain walls at electronic and ionic defects is the mechanism most likely responsible for this behaviour. At higher cycle numbers, a second mechanism sets in that changes the spatial distribution of the local electric field thus causing changes in the temporal behaviour of polarization switching in favour of slow, quasi-logarithmic dependence (see Fig. 7(a)). Calculated for these specimens the switching time distributions reveal the remarkably asymmetrical distributions in favour of larger switching times (see Fig. 8(a)). The onset of this asymmetry coincides with the appearance of cracks in the sample.[11] Therefore, micro-cracking seems the most plausible cause for the second fatigue mechanism. However, it has to be assumed that the first mechanism continues to be effective in the high cycle number regime as well. The essential feature of a crack is that annealing at an elevated temperature does not heal it, such that it remains in the material and influences any following switching event. This is the main reason why the specimens with higher cycle numbers still reveal the substantially asymmetric switching time spectra after annealing (see Fig. 8(b)), even though a partial restoration of the main switching characteristics occurs as depicted in Fig. 11(d)-(f).

For the Cu- stabilized 94BNT-6BT ferroelectric ceramic the present data show only small changes in the switching dynamics and maximum polarization even up to $10^7$ bipolar electric cycles (Figs. 9, 10 and 11(g)-(i)), which is in good agreement with previous fatigue studies.[57] The stability of the properties in this cycling regime sets this composition apart from the PZT- based materials discussed above, which show obvious degradation of switching dynamics when subjected to $10^7$ cycles of electric field. However, this advantage compared to the PZT- based PIC 151 composition quickly vanishes during the next decade of electric cycles. The switching dynamics and maximum polarization get strongly supressed and, in contrast to the other compositions, cannot be rejuvenated during a high temperature treatment.

Among the material classes discussed in the present manuscript, the system 94BNT- 6BT+Cu presents a special case as it belongs to the group of relaxor ferroelectrics. Earlier research has revealed the presence of a pseudo-cubic crystal structure for electrically unpoled 94BNT-6BT+Cu. Upon first field application the material undergoes a phase transformation into a polar state with tetragonal crystal symmetry that remains stable for least up to 100 bipolar cycles.[57] The importance of the stability of this field-induced polar phase on the properties of relaxor ferroelectrics has been widely discussed.[31,33,54,65]

Electric fatigue studies on the related relaxor-ferroelectric compositions $94(Bi_{1/2}Na_{1/2})TiO_3$-$6BaTiO_3$[58] and $[(Bi_{1/2}Na_{1/2})_{0.95}Ba_{0.05}]_{0.98}La_{0.02}TiO_3$[59] have revealed that bipolar electric cycling has a strong impact on the domain morphology of these systems. While long-range ferroelectric domains are created upon first electric field application due to the field-induced transition from the relaxor into the ferroelectric state, further electric cycling leads to a subsequent fragmentation of the domains even down to the nanoscale. The fragmentation process is understood to be initiated through the pinning of charged domain walls created during the polarization reversal process. The charged entities attract charge carriers in their surroundings, which agglomerate and stabilize the domain walls against further movement. For the material to be able to further compensate the cyclic external field, new domains have to be nucleated, leading to a subsequent increase of the domain wall density and a reduction of the average domain size.

The probability to nucleate a new domain depends among other factors on the crystal lattice distortion. Lower lattice distortion is associated with less elastic energy of the domain walls making it easier to introduce new domain walls into the system.[66] This can explain why domain fragmentation is pronounced for, *e.g.*, 94($Bi_{1/2}Na_{1/2}$)$TiO_3$-6$BaTiO_3$, showing significantly lower lattice distortion compared to, *e.g.,* the soft commercial PZT of the present study.[67]

The process of domain fragmentation not only allows for the progressive reduction of the local depolarization fields even though more and more domain walls become immobile. It helps as well to delay the development of internal stresses, which are discussed as source of mechanical degradation, such as crack formation, during bipolar electric cycling.[31,32] As such, the domain fragmentation mechanism can be seen as the underlying cause for the stable polarization values and switching dynamics during the first $10^7$ cycles for 94BNT-6BT+Cu in contrast to both PZT- based compositions. Furthermore, the dominance of this mechanism can even explain the sudden and irreversible property degradation of 94BNT-6BT+Cu during further electric field cycling. As mentioned above the fragmentation of the domains is associated with a constantly increasing domain wall density. While the presence of many domain walls allow the system to easily compensate depolarization electric fields and internal stresses, this assumption only holds up to a certain threshold. If too many domain walls are present they start impeding each other in their movement leading to degradation of the properties as, *e.g.*, discussed in the context of materials aging.[68,69] The onset of this so called "self-clamping" effect can be located to occur between $10^7$ and $10^8$ cycles for 94BNT-6BT+Cu. As the compensation mechanism falls short, a strong build-up of internal stresses and associated development of mechanical degradation is expected to occur in this cycling regime, explaining the sudden and irreversible degradation of the switching dynamics.

## 8. Conclusions

Polarization switching dynamics in ferroelectric ceramics is heavily influenced by fatigue. However, since fatigue mechanisms in different ceramics exhibit distinct features, the evolution of polarization reversal properties with fatigue also reveal different traits. These differences can be quantified when observing, on the one hand, macroscopic characteristics such as statistical switching time distributions and, on the other hand, microscopic characteristics such as activation fields of the local polarization reversal. Both kinds of characteristics can be obtained by analysis in terms of the Inhomogeneous Field Mechanism model. To adequately describe switching time distributions in fatigued ferroelectrics, the IFM - model was modified to include the presence of unswitchable or nonpolarizable regions in the material. In the modified form, the model is now able to capture polarization switching behaviour in composite materials consisting of switchable and unswitchable components, particularly, porous ferroelectric ceramics. In this paper, three different ceramics were exemplarily studied, Lanthanum doped PZT, commercial PZT material PIC 151 and Cu- stabilized 94BNT- 6BT.

Considering the fatigue behaviour of the PZT+La ceramic the charged defects suggest themselves as a most probable source of (reversible) evolution of the switching time distribution. Space charge accumulation as well as polar defect re-orientation affect first of all the non-180° switching events leading to a heavily fatigued state which, however, can be almost completely rejuvenated by thermal treatment. Commercial PIC 151 ceramic exhibits much better fatigue resistance and only weak changes in the switching spectrum up to the cycling number of $10^7$. By more electric cycling events, however, micro-cracking sets in presumably due to strains related to the non-180° switching processes. This leads to the fatigued state characterized by a very dispersive switching time spectrum which cannot be rejuvenated by annealing. Switching behaviour of the 94BNT-6BT+Cu ceramic demonstrates a very unusual two-stage evolution scenario in the course of fatigue due to specific structure properties of this material. Its most characteristic feature is progressive nano-fragmentation of polarization domains which prevents development of both high strains and high depolarization fields. This results in high fatigue resistance up to $10^7$ electric cycles followed by a sudden fatigue development by further cycling due to presumably "self-clamping" at high domain wall density.

## Acknowledgments

This work was supported by the Deutsche Forschungsgemeinschaft (DFG) within the collaborative research center SFB 595 (Electrical Fatigue of Functional Materials) and grants SE 941/17-1 and GE 1171/7-1. JG gratefully acknowledges support from the EU call H2020-MSCA-IF-2014 under project number 655866.